\documentclass[prb,preprintnumbers,amsmath,amssymb,floatfix]{revtex4}

\usepackage{graphicx}
\usepackage{subfigure}
\usepackage{dcolumn}
\usepackage{color}

\begin{document}

\title{The quantum tunnel effect of photon in one-dimensional photonic crystals}
\author{Ming-li Ren$^{a}$, Han Liu$^{a}$, Qing-Pan$^{a}$, Meng Han$^{a}$, Shan-Shan Tan$^{a}$, Zhang-Hui Liu$^{a}$ and Xiang-Yao Wu$^{a}$ \footnote{E-mail: wuxy65@126.com}}
 \affiliation{a. Institute of Physics, Jilin Normal
University, Siping 136000 China\\}

\begin{abstract}
In the paper, we have given the quantum transmissivity,
probability density and probability current density of photon in
one-dimensional photonic crystals $(AB)^N$ with the quantum theory
approach. We find the quantum transmissivity is identical to the
classical transmissivity. When the incident angle $\theta$ and
periodic number $N$ change the probability density and probability
current density are approximate periodic change, and their
amplitude are increased with the incident angles $\theta$ and
periodic number $N$ increasing. Otherwise, we find when the
frequency of incident photon is corresponding to transmissivity
$T=1$, the amplitude of the probability density is the largest.
When the frequency of incident photon is corresponding to
transmissivity $T=0$, the amplitude of the probability density
attenuate rapidly to zero, it indicates that there is the quantum
tunnel effect of photon in photonic crystals.

\vskip 5pt


\end{abstract}

\maketitle

\vskip 8pt
 {\bf 1. Introduction} \vskip 8pt
In 1987, E. Yablonovitch and S. John had pointed out the spread
behavior of photons in the periodical dielectric constant, and
termed such material Photonic Crystal [1, 2], which important
characteristics are: photon band gap, defect states, light
localization and so on. These characteristics may be used to
manufacture some high performance devices, such as pohotonic
crystal fiber [3], High performance mirror [4], optical filters
[5], Photonic crystal microcavity [6], splitters and combiners [7,
8]. optical limiters and amplifiers [9-12].

In the numerical calculations, there are many theory methods for
PCs, such as: the transfer matrix method (TMM) [13], the
finite-difference time-domain method (FDTD) [14], the plane-wave
expansion method (PWE) [15], the Green's function method [16], the
finite element method (FE) [17], the scattering matrix method [18]
and so on. These methods are form the classical electromagnetism
theory, i.e., classical method of light. Obviously, the quantum
theory of PCs should be necessary. In Ref. [19], we have proposed
quantum theory method of photon to study one-dimensional photonic
crystals transmissison characteristic, and calculated the quantum
dispersion relation, quantum transmissivity and reflectivity.  On
this basis, we have further studied the probability density and
probability current density of photon in one-dimensional photonic
crystals, and studied the effect of incident angle, refractive
index, period number, and frequency of incident photon on the
probability density and probability current density of photon in
one-dimensional photonic crystals. We find the quantum
transmissivity is identical to the classical transmissivity. When
the incident angle $\theta$ and periodic number $N$ change the
probability density and probability current density are
approximate periodic change, and their amplitude are increased
with the incident angles $\theta$ and periodic number $N$
increasing. Otherwise, we find when the frequency of incident
photon is corresponding to transmissivity $T=1$, the amplitude of
the probability density is the largest. When the frequency of
incident photon is corresponding to transmissivity $T=0$, the
amplitude of the probability density attenuate rapidly to zero, it
indicates that there is the quantum tunnel effect of photon in
photonic crystals.

\vskip 8pt

\newpage

\vskip 8pt
 {\bf 2. The probability density and
the probability current density of photon in one-dimensional
photonic crystals} \vskip 8pt

In the following, we shall calculate the probability density and
the probability current density of photon at every medium layer,
they are [19]

(1) When $0<x<a$, the photon wave function of medium layer $A$ at
the first period is
\begin{eqnarray} \psi_{A}^{1}(x)=\left(
\begin{array}
[c]{c}%
A_{kA}^{1}(1)\\
A_{kA}^{1}(2)\\
A_{kA}^{1}(3)%
\end{array}
\right)e^{i(K_{0}C_2\cdot x+K_{0}\sin\theta\cdot y)}
 + \left(
\begin{array}
[c]{c}%
A_{-kA}^{1}(1)\\
A_{-kA}^{1}(2)\\
A_{-kA}^{1}(3)%
\end{array}
\right)e^{i(-K_{0}C_2\cdot x+K_{0}\sin\theta\cdot y)},
\end{eqnarray}
the $\psi_{A}^{1}$ hermitian conjugate is
\begin{eqnarray}
{\psi_{A}^{1}}^{+}(x)&=&({A_{kA}^{1*}(1)}\hspace{0.08in}
{A_{kA}^{2*}(1)}\hspace{0.08in}
{A_{kA}^{3*}(1)})e^{-i(K_{0}C_2\cdot x+K_{0}\sin\theta\cdot
y)}\nonumber\\&+&({A_{-kA}^{1*}(1)}\hspace{0.08in}
{A_{-kA}^{2*}(1)}\hspace{0.08in}
{A_{-kA}^{3*}(1)})e^{-i(-K_{0}C_2\cdot x+K_{0}\sin\theta\cdot y)}.
\end{eqnarray}
With Eqs. (8) in the Ref. [19], we can calculate the probability
density of photon in one-dimensional photonic crystals. The
probability density of medium layer $A$ at the first period is
\begin{eqnarray}
\rho(x)={\psi_{A}^{1}}^{+}\cdot\psi_{A}^{1}&=&(|A_{kA}^{1}(1)|^2+|A_{kA}^{1}(2)|^2+
|A_{kA}^{1}(3)|^2)+(|A_{-kA}^{1}(1)|^2+|A_{-kA}^{1}(2)|^2+|A_{-kA}^{1}(3)|^2)
\nonumber\\&+&({A_{kA}^{1*}(1)}\cdot
A_{-kA}^{1}(1)+{A_{kA}^{1*}(2)}\cdot
A_{-kA}^{1}(2)+{A_{kA}^{1*}(3)}\cdot
A_{-kA}^{1}(3))e^{-2iK_{0}C_2\cdot
x}\nonumber\\&+&({A_{-kA}^{1*}(1)}\cdot
A_{kA}^{1}(1)+{A_{-kA}^{1*}(2)}\cdot
A_{kA}^{1}(2)+{A_{-kA}^{1*}(3)}\cdot
A_{kA}^{1}(3))e^{2iK_{0}C_2\cdot x}.
\end{eqnarray}

(2) When $a<x<a+b$, the photon wave function of medium layer $B$
at the first period is
\begin{eqnarray} \psi_{B}^{1}(x)=\left(
\begin{array}
[c]{c}%
B_{kB}^{1}(1)\\
B_{kB}^{1}(2)\\
B_{kB}^{1}(3)%
\end{array}
\right)e^{i(K_{0}C_3\cdot (x-a)+K_{0}\sin\theta\cdot y)}
 + \left(
\begin{array}
[c]{c}%
B_{-kB}^{1}(1)\\
B_{-kB}^{1}(2)\\
B_{-kB}^{1}(3)%
\end{array}
\right)e^{i(-K_{0}C_3\cdot (x-a)+K_{0}\sin\theta\cdot y)},
\end{eqnarray}
the $\psi_{B}^{1}$ hermitian conjugate is
\begin{eqnarray}
{\psi_{B}^{1}}^{+}(x)&=&({B_{kB}^{1*}(1)}\hspace{0.08in}
{B_{kB}^{2*}(1)}\hspace{0.08in}
{B_{kB}^{3*}(1)})e^{-i(K_{0}C_2\cdot (x-a)+K_{0}\sin\theta\cdot
y)}\nonumber\\&+&({B_{-kB}^{1*}(1)}\hspace{0.08in}
{B_{-kB}^{2*}(1)}\hspace{0.08in}
{B_{-kB}^{3*}(1)})e^{-i(-K_{0}C_2\cdot (x-a)+K_{0}\sin\theta\cdot
y)}
\end{eqnarray}
and probability density of medium layer $B$ at the first period is
\begin{eqnarray}
\rho(x)={\psi_{B}^{1}}^{+}\cdot\psi_{B}^{1}&=&(|B_{kB}^{1}(1)|^2+|B_{kB}^{1}(2)|^2+
|B_{kB}^{1}(3)|^2)+(|B_{-kB}^{1}(1)|^2+|B_{-kB}^{1}(2)|^2+|B_{-kB}^{1}(3)|^2)
\nonumber\\&+&({B_{kB}^{1*}(1)}\cdot
B_{-kB}^{1}(1)+{B_{kB}^{1*}(2)}\cdot
B_{-kB}^{1}(2)+{B_{kB}^{1*}(3)}\cdot
B_{-kB}^{1}(3))e^{-2iK_{0}C_2\cdot
(x-a)}\nonumber\\&+&({B_{-kB}^{1*}(1)}\cdot
B_{kB}^{1}(1)+{B_{-kB}^{1*}(2)}\cdot
B_{kB}^{1}(2)+{B_{-kB}^{1*}(3)}\cdot
B_{kB}^{1}(3))e^{2iK_{0}C_2\cdot (x-a)}.
\end{eqnarray}

(3) And accordingly, the photon wave function of medium layer $A$
at the $N-th$ period is
\begin{eqnarray} \psi_{A}^{N}=\left(
\begin{array}
[c]{c}%
A_{kA}^{N}(1)\\
A_{kA}^{N}(2)\\
A_{kA}^{N}(3)%
\end{array}
\right)e^{i(K_{0}C_2\cdot (x-(N-1)(a+b))+K_{0}\sin\theta\cdot y)}
 + \left(
\begin{array}
[c]{c}%
A_{-kA}^{N}(1)\\
A_{-kA}^{N}(2)\\
A_{-kA}^{N}(3)%
\end{array}
\right)e^{i(-K_{0}C_2\cdot (x-(N-1)(a+b))+K_{0}\sin\theta\cdot
y)},
\end{eqnarray}
the $\psi_{A}^{N}$ hermitian conjugate is
\begin{eqnarray}
{\psi_{A}^{N}}^{+}&=&({A_{kA}^{N*}(1)}\hspace{0.08in}
{A_{kA}^{N*}(2)}\hspace{0.08in}
{A_{kA}^{N*}(3)})e^{-i(K_{0}C_2\cdot
(x-(N-1)(a+b))+K_{0}\sin\theta\cdot
y)}\nonumber\\&+&({A_{-kA}^{N*}(1)}\hspace{0.08in}
{A_{-kA}^{N*}(2)}\hspace{0.08in}
{A_{-kA}^{N*}(3)})e^{-i(-K_{0}C_2\cdot
(x-(N-1)(a+b))+K_{0}\sin\theta\cdot y)},
\end{eqnarray}
and the probability density of medium layer $A$ at the $N-th$
period is
\begin{eqnarray}
\rho(x)={\psi_{A}^{N}}^{+}\cdot\psi_{A}^{N}&=&(|A_{kA}^{N}(1)|^2+|A_{kA}^{N}(2)|^2+
|A_{kA}^{N}(3)|^2)+(|A_{-kA}^{N}(1)|^2+|A_{-kA}^{N}(2)|^2+|A_{-kA}^{N}(3)|^2)
\nonumber\\&+&({A_{kA}^{N*}(1)}\cdot
A_{-kA}^{N}(1)+{A_{kA}^{N*}(2)}\cdot
A_{-kA}^{N}(2)+{A_{kA}^{N*}(3)}\cdot
A_{-kA}^{N}(3))e^{-2iK_{0}C_2\cdot
(x-(N-1)(a+b))}\nonumber\\&+&({A_{-kA}^{N*}(1)}\cdot
A_{kA}^{N}(1)+{A_{-kA}^{N*}(2)}\cdot
A_{kA}^{N}(2)+{A_{-kA}^{N*}(3)}\cdot
A_{kA}^{N}(3))e^{2iK_{0}C_2\cdot (x-(N-1)(a+b))},
\end{eqnarray}
where $K_0=\frac{\omega}{c}$, $\omega$ is the angular frequency of
incident photon, $c$ is the velocity of light in vacuum,
$C_1=\sqrt{1-\sin^{2}\theta}$,
$C_2=\sqrt{n_A^{2}-\sin^{2}\theta}$,
$C_3=\sqrt{n_B^{2}-\sin^{2}\theta}$, $\theta$ is the incident
angle, and the parameters $a$ and $b$ are the thickness of media
$A$ and $B$, respectively.

(4) The photon wave function of medium layer $B$ at the $N-th$
period is
\begin{eqnarray} \psi_{B}^{N}=\left(
\begin{array}
[c]{c}%
B_{kB}^{N}(1)\\
B_{kB}^{N}(2)\\
B_{kB}^{N}(3)%
\end{array}
\right)e^{i(K_{0}C_3\cdot (x-(Na+(N-1)b))+K_{0}\sin\theta\cdot y)}
 + \left(
\begin{array}
[c]{c}%
B_{-kB}^{N}(1)\\
B_{-kB}^{N}(2)\\
B_{-kB}^{N}(3)%
\end{array}
\right)e^{i(-K_{0}C_3\cdot (x-(Na+(N-1)b))+K_{0}\sin\theta\cdot
y)},
\end{eqnarray}
the $\psi_{B}^{N}$ hermitian conjugate is
\begin{eqnarray}
{\psi_{B}^{N}}^{+}&=&({B_{kB}^{N*}(1)}\hspace{0.08in}
{B_{kB}^{N*}(2)}\hspace{0.08in}
{B_{kB}^{N*}(3)})e^{-i(K_{0}C_2\cdot
(x-(Na+(N-1)b))+K_{0}\sin\theta\cdot
y)}\nonumber\\&+&({B_{-kB}^{N*}(1)}\hspace{0.08in}
{B_{-kB}^{N*}(2)}\hspace{0.08in}
{B_{-kB}^{N*}(3)})e^{-i(-K_{0}C_2\cdot
(x-(Na+(N-1)b))+K_{0}\sin\theta\cdot y)},
\end{eqnarray}
and the probability density of medium layer $B$ at the $N-th$
\begin{eqnarray}
\rho(x)={\psi_{B}^{N}}^{+}\cdot\psi_{B}^{N}&=&(|B_{kB}^{N}(1)|^2+|B_{kB}^{N}(2)|^2+
|B_{kB}^{N}(3)|^2)+(|B_{-kB}^{N}(1)|^2+|B_{-kB}^{N}(2)|^2+|B_{-kB}^{N}(3)|^2)
\nonumber\\&+&({B_{kB}^{N*}(1)}\cdot
B_{-kB}^{N}(1)+{B_{kB}^{N*}(2)}\cdot
B_{-kB}^{N}(2)+{B_{kB}^{N*}(3)}\cdot
B_{-kB}^{N}(3))e^{-2iK_{0}C_2\cdot
(x-(Na+(N-1)b))}\nonumber\\&+&({B_{-kB}^{N*}(1)}\cdot
B_{kB}^{N}(1)+{B_{-kB}^{N*}(2)}\cdot
B_{kB}^{N}(2)+{B_{-kB}^{N*}(3)}\cdot
B_{kB}^{N}(3))e^{2iK_{0}C_2\cdot (x-(Na+(N-1)b))},
\end{eqnarray}
Where $Na+(N-1)b<x<N(a+b)$.

With Eqs. (9) in the Ref. [19], we can obtain the probability
current density of photon at every medium layer, they are

(1) when $0<x<a$, the probability current density of medium layer
$A$ at the first period is
\begin{eqnarray}
J_{x}=c{\psi_{A}^{1}}^{+}\alpha_{x}\psi_{A}^{1}&=&c[(iA_{kA}^{1*}(3)\cdot
A_{kA}^{1}(2)-iA_{kA}^{1*}(2)\cdot
A_{kA}^{1}(3))\nonumber\\&+&(iA_{kA}^{1*}(3)\cdot
A_{-kA}^{1}(2)-iA_{kA}^{1*}(2)\cdot
A_{-kA}^{1}(3))e^{-2iK_{0}C_2\cdot x}\nonumber\\&+&
(iA_{-kA}^{1*}(3)\cdot A_{kA}^{1}(2)-iA_{-kA}^{1*}(2)\cdot
A_{kA}^{1}(3))e^{2iK_{0}C_2\cdot x}\nonumber\\&+&
(iA_{-kA}^{1*}(3)\cdot A_{-kA}^{1}(2)-iA_{-kA}^{1*}(2)\cdot
A_{-kA}^{1}(3))].
\end{eqnarray}

(2) when $a<x<a+b$, the probability current density of medium
layer $B$ at the first period is
\begin{eqnarray}
J_{x}=c{\psi_{B}^{1}}^{+}\alpha_{x}\psi_{B}^{1}&=&c[(iB_{kB}^{1*}(3)\cdot
B_{kB}^{1}(2)-iB_{kB}^{1*}(2)\cdot
B_{kB}^{1}(3))\nonumber\\&+&(iB_{kB}^{1*}(3)\cdot
B_{-kB}^{1}(2)-iB_{kB}^{1*}(2)\cdot
B_{-kB}^{1}(3))e^{-2iK_{0}C_3\cdot (x-a)}\nonumber\\&+&
(iB_{-kB}^{1*}(3)\cdot B_{kB}^{1}(2)-iB_{-kB}^{1*}(2)\cdot
B_{kB}^{1}(3))e^{2iK_{0}C_3\cdot (x-a)}\nonumber\\&+&
(iB_{-kB}^{1*}(3)\cdot B_{-kB}^{1}(2)-iB_{-kB}^{1*}(2)\cdot
B_{-kB}^{1}(3))].
\end{eqnarray}

(3) The probability current density of medium layer $A$ at the
$N-th$ period is
\begin{eqnarray}
J_{x}=c{\psi_{A}^{N}}^{+}\alpha_{x}\psi_{A}^{N}&=&c[(iA_{kA}^{N*}(3)\cdot
A_{kA}^{N}(2)-iA_{kA}^{N*}(2)\cdot
A_{kA}^{N}(3))\nonumber\\&+&(iA_{kA}^{N*}(3)\cdot
A_{-kA}^{N}(2)-iA_{kA}^{N*}(2)\cdot
A_{-kA}^{N}(3))e^{-2iK_{0}C_2\cdot (x-(N-1)(a+b))}\nonumber\\&+&
(iA_{-kA}^{N*}(3)\cdot A_{kA}^{N}(2)-iA_{-kA}^{N*}(2)\cdot
A_{kA}^{N}(3))e^{2iK_{0}C_2\cdot (x-(N-1)(a+b))}\nonumber\\&+&
(iA_{-kA}^{N*}(3)\cdot A_{-kA}^{N}(2)-iA_{-kA}^{N*}(2)\cdot
A_{-kA}^{N}(3))],
\end{eqnarray}

where $(N-1)(a+b)<x<Na+(N-1)b$.

(4) The probability current density of medium layer $B$ at the
$N-th$ period is
\begin{eqnarray}
J(x)=c{\psi_{B}^{N}}^{+}\alpha_{x}\psi_{B}^{N}&=&c[(iB_{kB}^{N*}(3)\cdot
B_{kB}^{N}(2)-iB_{kB}^{N*}(2)\cdot
B_{kB}^{N}(3))\nonumber\\&+&(iB_{kB}^{N*}(3)\cdot
B_{-kB}^{N}(2)-iB_{kB}^{N*}(2)\cdot
B_{-kB}^{N}(3))e^{-2iK_{0}C_3\cdot (x-(Na+(N-1)b))}\nonumber\\&+&
(iB_{-kB}^{N*}(3)\cdot B_{kB}^{N}(2)-iB_{-kB}^{1N*}(2)\cdot
B_{kB}^{N}(3))e^{2iK_{0}C_3\cdot (x-(Na+(N-1)b))}\nonumber\\&+&
(iB_{-kB}^{N*}(3)\cdot B_{-kB}^{N}(2)-iB_{-kB}^{N*}(2)\cdot
B_{-kB}^{N}(3))],
\end{eqnarray}

where $Na+(N-1)b<x<N(a+b)$.

With the continuation of wave function and its derivative, at the
$j-th$ period ,we can give the relation between $A_{kA}^{j}(i)
(i=(1, 2, 3))$, $A_{-kA}^{j}(i) (i=(1, 2, 3))$ and $F_{i} (i=(1,
2, 3, 4, 5, 6))$, and $B_{kA}^{j}(i) (i=(1, 2, 3))$,
$B_{-kA}^{j}(i) (i=(1, 2, 3))$ and $F_{i} (i=(1, 2, 3, 4, 5, 6))$,
they are [19]

\begin{eqnarray}
\left ( \begin{array}{cccccc}
A_{kA}^{j}(1)\\
A_{kA}^{j}(2)\\
A_{kA}^{j}(3)\\
A_{-kA}^{j}(1)\\
A_{-kA}^{j}(2)\\
A_{-kA}^{j}(3)\\
   \end{array}
   \right )=(M_{B}\cdot M_{A})^{j-1}\cdot M_{0}\left (
\begin{array}{cccccc}
F_{1}\\
F_{2}\\
F_{3}\\
F_{4}\\
F_{5}\\
F_{6}\\
   \end{array}
\right).
\end{eqnarray}
and
\begin{eqnarray}
\left ( \begin{array}{cccccc}
B_{kB}^{j}(1)\\
B_{kB}^{j}(2)\\
B_{kB}^{j}(3)\\
B_{-kB}^{j}(1)\\
B_{-kB}^{j}(2)\\
B_{-kB}^{j}(3)\\
   \end{array}
   \right )=(M_{A}\cdot M_{B})^{j-1}\cdot M_{A}\cdot M_{0}\left (
\begin{array}{cccccc}
F_{1}\\
F_{2}\\
F_{3}\\
F_{4}\\
F_{5}\\
F_{6}\\
   \end{array}
\right).
\end{eqnarray}

\begin{eqnarray}M_{0}=\frac{1}{2} \left (
\begin{array}{cccccc}
p \hspace{0.1in} 0 \hspace{0.1in} 0 \hspace{0.1in} q \hspace{0.1in} 0 \hspace{0.1in} 0  \\
0 \hspace{0.1in} p \hspace{0.1in} 0 \hspace{0.1in} 0 \hspace{0.1in} q \hspace{0.1in} 0  \\
0 \hspace{0.1in} 0 \hspace{0.1in} p \hspace{0.1in} 0 \hspace{0.1in} 0 \hspace{0.1in} q \\
q \hspace{0.1in} 0 \hspace{0.1in} 0 \hspace{0.1in} p \hspace{0.1in} 0 \hspace{0.1in} 0  \\
0 \hspace{0.1in} q \hspace{0.1in} 0 \hspace{0.1in} 0 \hspace{0.1in} p \hspace{0.1in} 0  \\
0 \hspace{0.1in} 0 \hspace{0.1in} q \hspace{0.1in} 0 \hspace{0.1in} 0 \hspace{0.1in} p \\
\end{array}
   \right ).
\end{eqnarray}

\begin{eqnarray}
M_{A}=\frac{1}{2} \left ( \begin{array}{cccccc}
g_{1} \hspace{0.1in} 0 \hspace{0.1in} 0 \hspace{0.1in} g_{2} \hspace{0.1in} 0 \hspace{0.1in} 0  \\
0 \hspace{0.1in} g_{1} \hspace{0.1in} 0 \hspace{0.1in} 0 \hspace{0.1in} g_{2} \hspace{0.1in} 0  \\
0 \hspace{0.1in} 0 \hspace{0.1in} g_{1} \hspace{0.1in} 0 \hspace{0.1in} 0 \hspace{0.1in} g_{2}  \\
g_{3} \hspace{0.1in} 0 \hspace{0.1in} 0 \hspace{0.1in} g_{4} \hspace{0.1in} 0 \hspace{0.1in} 0  \\
0 \hspace{0.1in} g_{3} \hspace{0.1in} 0 \hspace{0.1in} 0 \hspace{0.1in} g_{4} \hspace{0.1in} 0  \\
0 \hspace{0.1in} 0 \hspace{0.1in} g_{3} \hspace{0.1in} 0 \hspace{0.1in} 0 \hspace{0.1in} g_{4}  \\
\end{array}
   \right ).
\end{eqnarray}

\begin{eqnarray}
M_{B}=\frac{1}{2} \left ( \begin{array}{cccccc}
h_{1} \hspace{0.1in} 0 \hspace{0.1in} 0 \hspace{0.1in} h_{2} \hspace{0.1in} 0 \hspace{0.1in} 0  \\
0 \hspace{0.1in} h_{1} \hspace{0.1in} 0 \hspace{0.1in} 0 \hspace{0.1in} h_{2} \hspace{0.1in} 0  \\
0 \hspace{0.1in} 0 \hspace{0.1in} h_{1} \hspace{0.1in} 0 \hspace{0.1in} 0 \hspace{0.1in} h_{2}  \\
h_{3} \hspace{0.1in} 0 \hspace{0.1in} 0 \hspace{0.1in} h_{4} \hspace{0.1in} 0 \hspace{0.1in} 0  \\
0 \hspace{0.1in} h_{3} \hspace{0.1in} 0 \hspace{0.1in} 0 \hspace{0.1in} h_{4} \hspace{0.1in} 0  \\
0 \hspace{0.1in} 0 \hspace{0.1in} h_{3} \hspace{0.1in} 0 \hspace{0.1in} 0 \hspace{0.1in} h_{4}  \\
\end{array}
   \right ).
\end{eqnarray}

where $p=1+\frac{C_1}{C_2}$, $q=1-\frac{C_1}{C_2}$,
$g_{1}=(1+\frac{C_2}{C_3})e^{{iK_{0}C_2}\cdot a}$,
$g_{2}=(1-\frac{C_2}{C_3})e^{{-iK_{0}C_2}\cdot a}$,
$g_{3}=(1-\frac{C_2}{C_3})e^{{iK_{0}C_2}\cdot a}$,
$g_{4}=(1+\frac{C_2}{C_3})e^{{-iK_{0}C_2}\cdot a}$,
$h_{1}=(1+\frac{C_3}{C_2})e^{{iK_{0}C_3}\cdot b}$,
$h_{2}=(1-\frac{C_3}{C_2})e^{{-iK_{0}C_3}\cdot b}$,
$h_{3}=(1-\frac{C_3}{C_2})e^{{iK_{0}C_3}\cdot b}$,
$h_{4}=(1+\frac{C_3}{C_2})e^{{-iK_{0}C_3}\cdot b}$.

 \vskip 8pt {\bf 3. Numerical result} \vskip 8pt

In this section, we report our numerical results of quantum
transmissivity, the probability density, and the probability
current density of one-dimensional photonic crystals with
structure $(AB)^{N}$. The main parameters are as follows: the
refractive index of media $A$ and $B$ are $n_{a}=2.68$ and
$n_{b}=1.68$, their thickness are $a=200nm$ and $b=300nm$, the
period numbers $N=10$, the central frequency $\omega_{0}=171 THz$.
In numerical calculation, we compare quantum transmissivity with
classical transmissivity, and further study the effects of
different parameters and structures of one-dimensional photonic
crystals on the probability density and the probability current
density. In Ref. [19], with Eq. (88), we can calculate the quantum
transmissivity, which is shown in Fig. 1. The Fig. 1 (a) is the
classical transmissivity, and the Fig. 1 (b) is the quantum
transmissivity. Comparing Fig. 1 (a) with (b), we can find the
quantum transmissivity is identical to classical transmissivity.
In Figs. 2 and 3, we study the effect of incident angles $\theta$
on the the probability density and probability current density,
respectively. The Figs. 2 and 3 (a), (b) and (c) correspond to
incident angles $\theta=\frac{\pi}{4}$, $ \frac{\pi}{6}$, and
$\frac{\pi}{10}$, respectively. From Figs. 2 and 3, we can find
the probability density and probability current density are
approximate periodic change with the propagation distance $x$ of
photon in the photonic crystal. The wave peak, wave trough and
total wave amplitude of the probability density and probability
current density increase with the incident angles $\theta$
increasing, and the periodic number of the probability density are
almost unchanged. In Fig. 4, we study the effect of the periodic
number $N$ on the the probability density. The Fig. 5 (a), (b) and
(c) correspond to periodic number $N=8$, $9$ and $10$,
respectively. From Fig. 5, we can find the probability density are
approximate periodic change with the propagation distance $x$ of
photon in the photonic crystal, The wave peak, wave trough and
total wave amplitude of the probability density increase with the
periodic number $N$ increasing, and the periodic number of the
probability density is nearly invariable. In Fig. 5, we study the
effects of different photonic crystal structure on the probability
density. The Fig. 5 (a) and (b) correspond to the structure
$(AB)^{10}$ and $(AB)^{5}(BA)^{5}$ (mirror structure),
respectively. From Fig. 5, we can find the probability density of
mirror structure photonic crystal obviously reduce. In Fig. 6, we
further study the effect of the frequency of incident photon on
the the probability density. The Fig. 6 (a), (b) and (c)
correspond to the frequency of incident photon $\omega
=1.25\omega_0$, $1.5\omega_0$, and $3.2\omega_0$, which are the
frequencies of the points $A$, $B$ and $C$ at Fig. 1, their
transmissivity are $T=1$, $T=0.8$ and $T=0$, respectively. In Fig.
6 (a), we take the frequency of incident photon corresponding to
transmissivity $T=1$, the wave peak and total wave amplitude of
the probability density is the largest. In Fig. 6 (b), we take the
frequency of incident photon corresponding to transmissivity
$T=0.8$, the wave peak and total wave amplitude of the probability
density is smaller than the Fig. 6 (a). In Fig. 6 (c), we take the
frequency of incident photon corresponding to transmissivity
$T=0$, i.e, the photon frequency of forbidden band, the wave peak
and total wave amplitude of the probability density damp rapidly
to zero, it indicates there is the quantum tunnel effect of photon
in photonic crystal.

\vskip 8pt
 {\bf 4. Conclusions} \vskip 8pt

In the paper, we have given the quantum transmissivity,
probability density and probability current density of
one-dimensional photonic crystals $(AB)^N$ with the quantum theory
of photon. By calculation, we find the quantum transmissivity is
identical to the classical transmissivity. When the incident angle
$\theta$ and periodic number $N$ change the probability density
and probability current density are approximate periodic change
with the propagation distance $x$ of photon in the photonic
crystal, and the wave amplitudes of the probability density and
probability current density increase with the incident angles
$\theta$ and periodic number $N$ increasing. Otherwise, we find
the the frequency of incident photon has a certain effect on
probability density and probability current density. When the
frequency of incident photon is corresponding to transmissivity
$T=1$, the wave peak and total wave amplitude of the probability
density is the largest. When the frequency of incident photon is
corresponding to transmissivity $T=0$, the wave peak and total
wave amplitude of the probability density attenuate rapidly to
zero, it indicates there is the quantum tunnel effect of photon in
photonic crystals.

\vskip 5pt
 {\bf 5. Acknowledgment} \vskip 5pt
This work was supported by the Scientific and Technological
Development Foundation of Jilin Province (no.20130101031JC).

\begin{figure}[htbp]
\includegraphics[width=16cm, height=10cm]{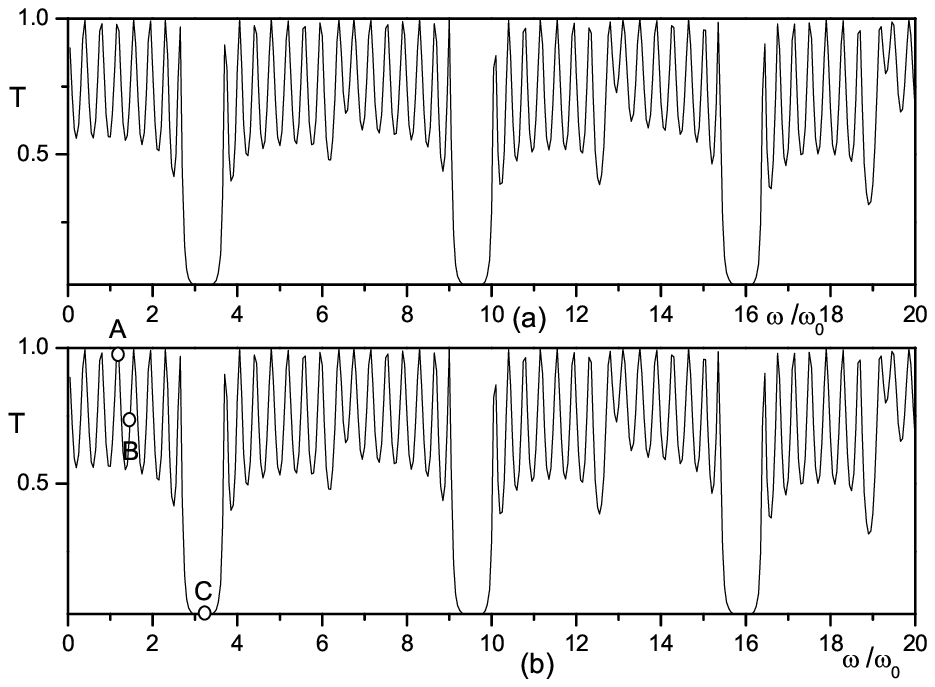}
\caption{The transmissivity of one-dimensional photonic crystals
$(AB)^{10}$. (a) the classical transmissivity, (b) the quantum
transmissivity.}
\end{figure}

\begin{figure}[htbp]
\includegraphics[width=16cm, height=8cm]{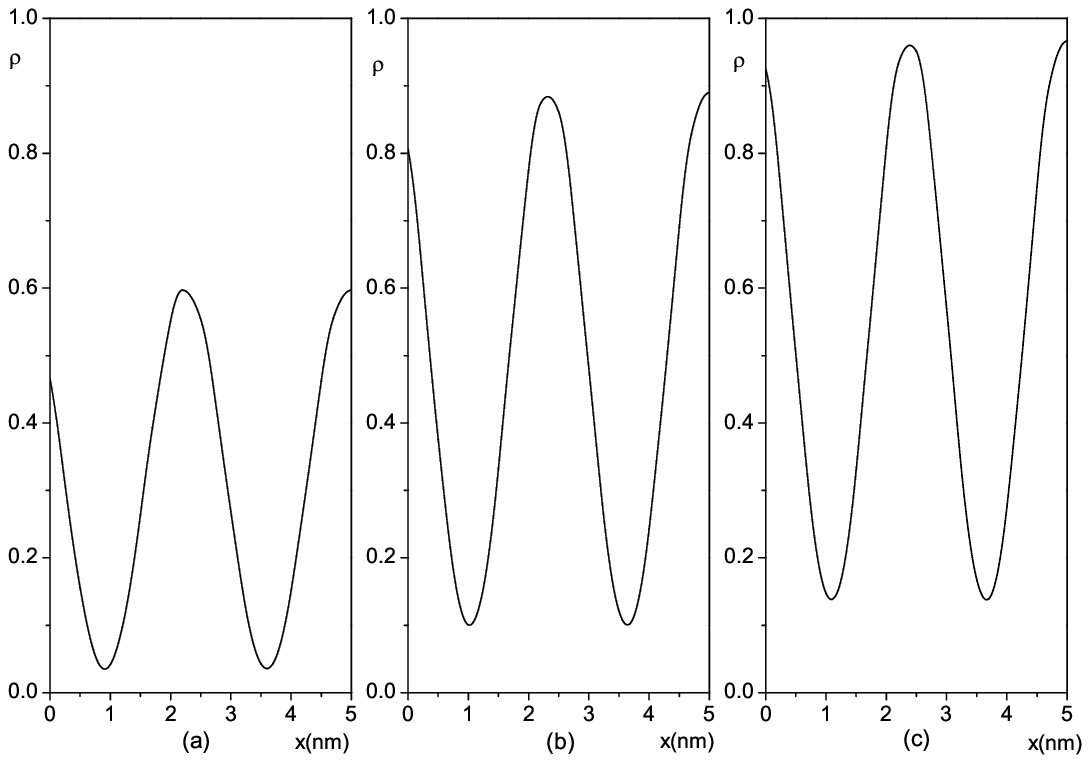}
\caption{The probability density of one-dimensional photonic
crystals $(AB)^{10}$ for different incident angle $\theta$. (a)
$\theta = \frac{\pi}{3}$, (b) $\theta = \frac{\pi}{4}$, (c)
$\theta = \frac{\pi}{5}$.}
\end{figure}

\begin{figure}[htbp]
\includegraphics[width=16cm, height=8cm]{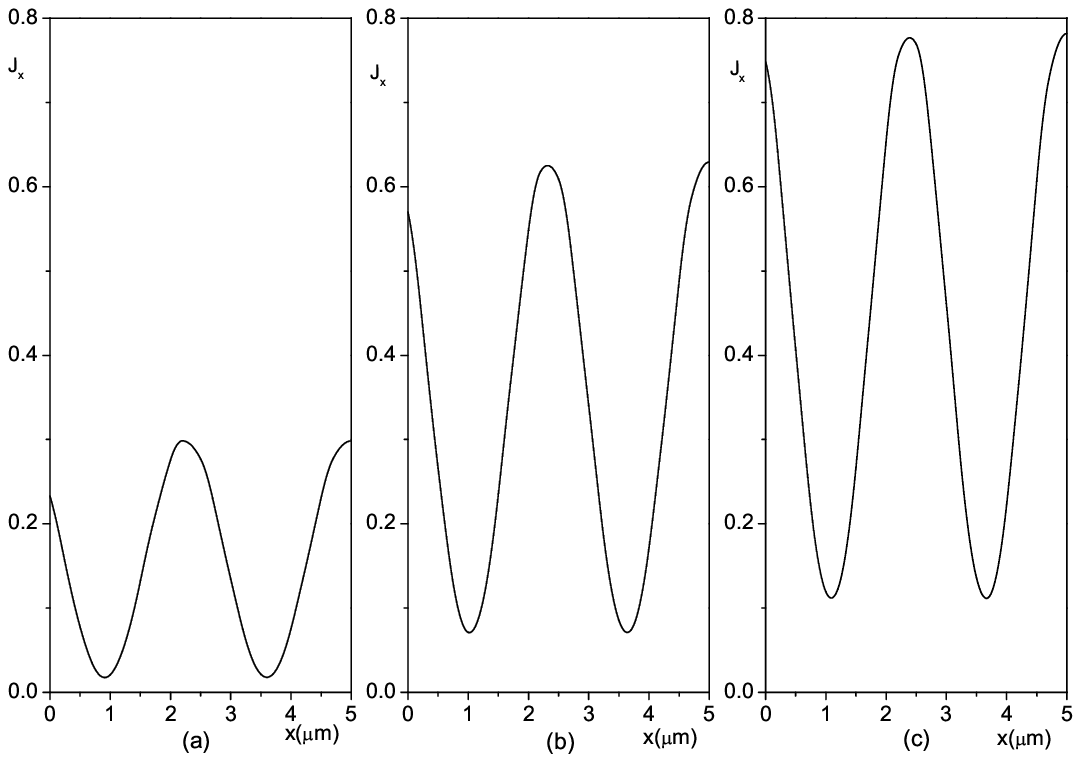}
\caption{The probability current density of one-dimensional
photonic crystals $(AB)^{10}$ for different incident angle
$\theta$. (a) $\theta = \frac{\pi}{3}$, (b) $\theta =
\frac{\pi}{4}$, (c) $\theta = \frac{\pi}{5}$.}
\end{figure}

\begin{figure}[htbp]
\includegraphics[width=16cm, height=8cm]{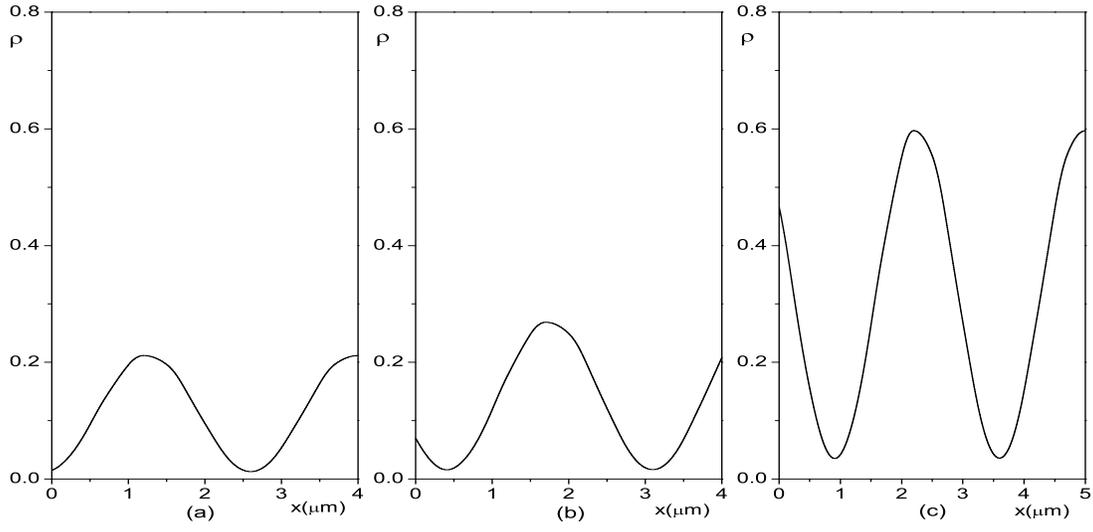}
\caption{The probability density of different one-dimensional
photonic crystals for different periodic number $N$. (a) $N=8$,
(b) $N=9$, (c) $N=10$.}
\end{figure}

\begin{figure}[htbp]
\includegraphics[width=16cm, height=8cm]{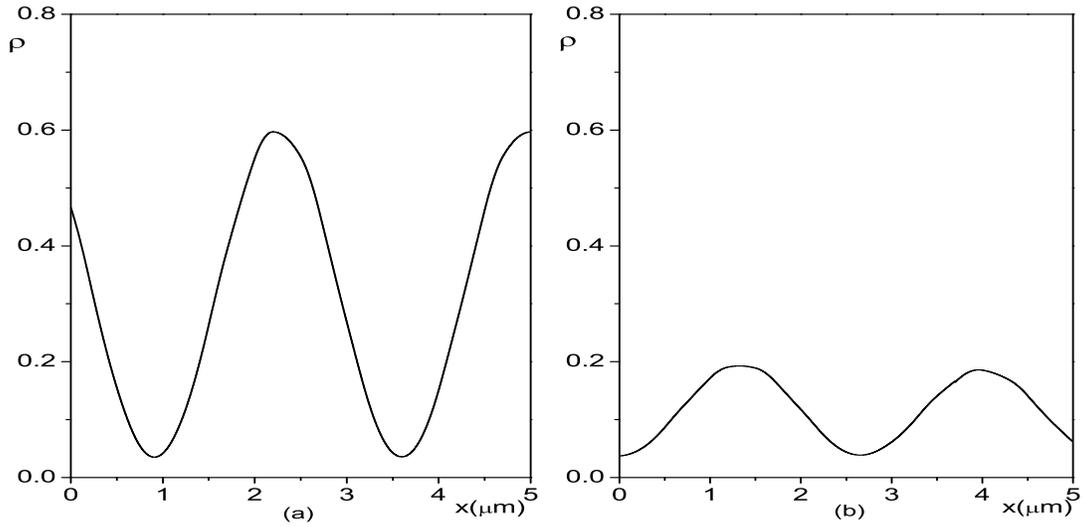}
\caption{The probability density of different one-dimensional
photonic crystals with different structure. (a) $(AB)^{10}$, (b)
$(AB)^{5}(BA)^{5}$.}
\end{figure}
\begin{figure}[htbp]
\includegraphics[width=16cm, height=8cm]{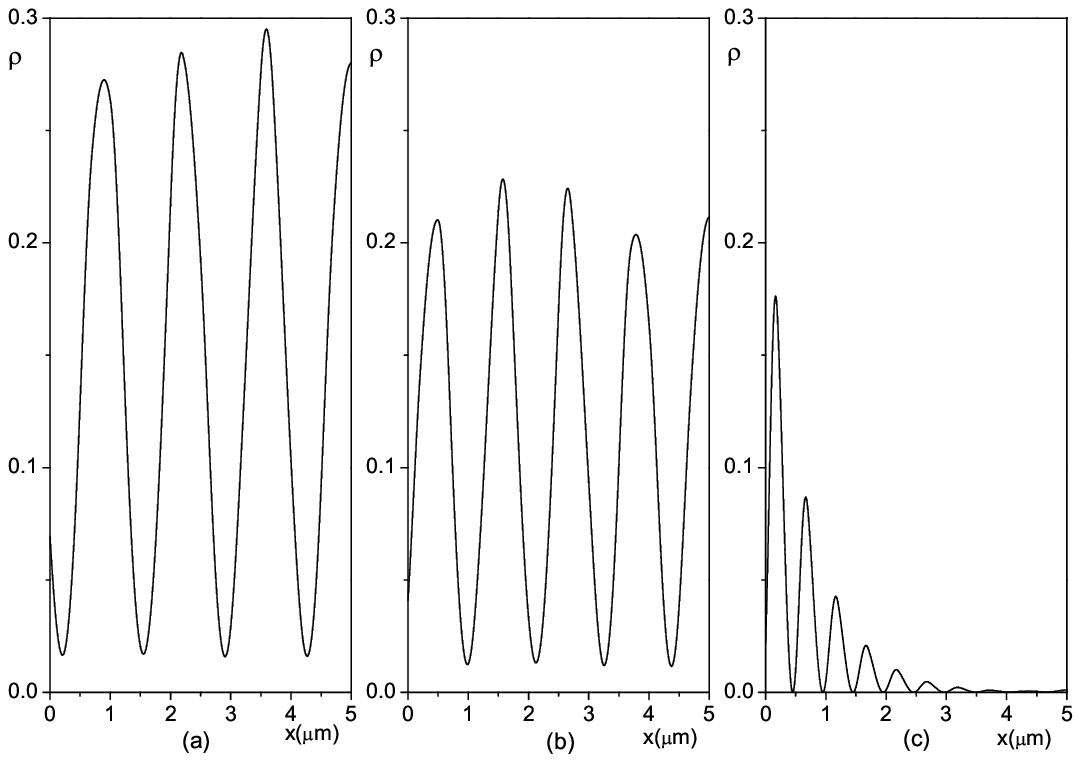}
\caption{The probability density of one-dimensional photonic
crystals $(AB)^{10}$ with different $\omega$. (a)
$\omega=1.25\omega_{0} \hspace{0.05in}(T=1)$, (b)
$\omega=1.5\omega_{0} \hspace{0.05in}(T=0.8)$, (c)
$\omega=3.2\omega_{0} \hspace{0.05in}(T=0)$.}
\end{figure}


\begin{thebibliography}{10}

\bibitem{s1}
J. D. Joannopoulos, P. R. Villeneuve, and S. Fan,  Nature {\bf386} 143 (1997).

\bibitem{s2}
P. Russell, Science {\bf 299} 358 (2003).

\bibitem{s3}
M. S. Habib£¬M. S. Habib£¬M. I. Hasan£¬et al, Optical Fiber Technology£¬{\bf20} 328 (2014).

\bibitem{s4}
Y. Fink£¬J. N. Winn£¬S. Fan,  Science£¬{\bf282} 1679 (1998).


\bibitem{s5}
Z. T. Ma£¬K. Ogusu, Optics Communications£¬{\bf284} 1192 (2011)£®


\bibitem{s6} A. H. Safavi-Naeini£¬ O. Painter, Optics Express£¬{\bf18} 14926 (2010).


\bibitem{s7}
S. John, Phys. Rev. Lett. {\bf 58} 2486 (1987).

\bibitem{s8}
V. S. C. Manga Rao and S. Hughes., Phys. Rev B {\bf 75} 205437
(2007).

\bibitem{s9}
M. L. M. Balistreri, H. Gersen, J. P. Korterik, L. Kuipers, and N. F. van Hulst, Science {\bf 294}, 1080¨C1082 (2001).

\bibitem{s10}
T. Lund-Hansen, S. Stobbe, B. Julsgaard, H. Thyrrestrup, T.S$\ddot{u}$nner, M. Kamp, A. Forchel, and P. Lodahl., Phys. Rev. Lett.
{\bf 101} 113903 (2008).

\bibitem{s11}
S. J. Dewhurst, D. Granados, D. J. P. Ellis, A. J. Bennett, R. B. Patel, I. Farrer, D. Anderson, G. A. C. Jones, D. A. Ritchie, and
A. J. Shields., Appl. Phys. Lett. {\bf 96} 031109 (2010).

\bibitem{s12}
P. A. Kalozoumis, G. Theocharis, V. Achilleos, S. F. elix, O. Richoux, and V. Pagneux, Phys. Rev. A 98, 023838 (2018).

\bibitem{s13}
S. G. Johnson and J. D. Joannopoulos, Optics Express {\bf 8}, no.3, 173 (2001).

\bibitem{s14}
E. G. Alivizatos, I. D. Chremmos, N. L. Tsitsas, et. al., J. Opt. Soc. Am. A {\bf 21(5)}, 847 (2004).


\bibitem{s15}
W. Zhu, X. Fang, D. Li, Y. Sun, Y. Li, Y. Jing, and H. Chen, Phys. Rev. Lett. 121, 124501 (2018).

\bibitem{s16}
K. S. Choi, H. Deng, J. Laurat, and H. J. Kimble, Nature {\bf
452}, 67 (2006).

\bibitem{s17}
M. C. Lin and R. F. Jao, Optics Express {\bf 15}, 207 (2007).

\bibitem{s18}
W. S. Mohammed, L. Vaissie and E. G. Johnson, Optical Engineering
{\bf 42(8)}, 2311 (2003).

\bibitem{s19}
Xiang-Yao Wu, Qing-Pan, Xiao-Ru Zhang, Han Liu, Fu-Quan Yang,
Ji-Ping Liu, Ji Ma, Hong-Chun Yuan, Physica E, {\bf 114} 113563
(2019).

\newpage
\end{thebibliography}
\end{document}